\documentclass[fleqn,usenatbib]{mnras}

\usepackage{newtxtext,newtxmath}
\usepackage{float}
\usepackage{ulem}
\usepackage{subfigure}
\usepackage[T1]{fontenc}

\DeclareRobustCommand{\VAN}[3]{#2}
\let\VANthebibliography\thebibliography
\def\thebibliography{\DeclareRobustCommand{\VAN}[3]{##3}\VANthebibliography}

\usepackage{graphicx}	
\usepackage{amsmath}	

\title[VSF from BOSS DR16]{Cosmological Constraints using the Void Size Function Data from BOSS DR16}

\author[Y. Song et al.]{
Yingxiao Song$^{1,2}$,
Yan Gong$^{1,2,3}$\thanks{Email:gongyan@bao.ac.cn},
Xingchen Zhou$^{1}$,
Haitao Miao$^{1}$,
Kwan Chuen Chan$^{4,5}$,\newauthor
and
Xuelei Chen$^{1,2,6,7}$
\\\\
$^{1}$National Astronomical Observatories, Chinese Academy of Sciences,20A Datun Road, Beijing 100012, China\\
$^{2}$School of Astronomy and Space Sciences, University of Chinese Academy of Sciences(UCAS),Yuquan Road NO.19A Beijing 100049, China\\
$^{3}$Science Center for China Space Station Telescope, National Astronomical Observatories, Chinese Academy of Sciences,\\20A Datun Road, Beijing 100101, China\\
$^{4}$School of Physics and Astronomy, Sun Yat-sen University, 2 Daxue Road, Tangjia, Zhuhai, 519082, China\\
$^{5}$CSST Science Center for the Guangdong-Hongkong-Macau Greater Bay Area, SYSU, Zhuhai, 519082, China\\
$^{6}$Department of Physics, College of Sciences, Northeastern University, Shenyang 110819, China\\
$^{7}$Centre for High Energy Physics, Peking University, Beijing 100871, China\\
}
\date{Accepted XXX. Received YYY; in original form ZZZ}

\pubyear{2015}

\begin{document}
\label{firstpage}
\pagerange{\pageref{firstpage}--\pageref{lastpage}}
\maketitle
 
\begin{abstract}
We measure the void size function (VSF) from the Baryon Oscillation Spectroscopic Survey (BOSS DR16) and perform the cosmological constraints. The BOSS DR16 galaxy sample is selected in the redshift range from $z = 0.2$ to 0.8, considering the selection criteria based on galaxy number density. We identify non-spherical voids from this galaxy catalog using the Voronoi tessellation and watershed algorithm without assuming any void shape. We select the void samples based on the void ellipticity, and derive the VSFs in two redshift bins, i.e. $z=0.2-0.5$ and $0.5-0.8$. The VSF model we use is based on the excursion-set theory, including the void linear underdensity threshold $\delta_{\rm v}$ and the redshift space distortion (RSD) parameter $B$. The Markov Chain Monte Carlo (MCMC) method is applied to perform the joint constraints on the cosmological and void parameters. 
We find that the VSF measurement from BOSS DR16 gives $w = -1.263_{-0.396}^{+0.329}$, $\Omega_{\rm m} = 0.293_{-0.053}^{+0.060}$, and $\sigma_8 = 0.897_{-0.192}^{+0.159}$, 
which can be a good complementary probe to galaxy clustering measurements. Our method demonstrates the potential of using the VSF to study cosmological models, and it can provide a reference for the future VSF analysis in the upcoming galaxy spectroscopic surveys.

\end{abstract}


\section {introduction} \label{sec:intro}
In recent decades, our understanding of the cosmic large-scale structure (LSS), which describes the distribution of matter on cosmic scales, has advanced significantly. This progress has been primarily driven by extensive and high-precision galaxy surveys. As these surveys expand, they provide larger galaxy samples and increased survey volumes, which make the features of the LSS more recognizable. Among these features, cosmic voids, i.e. regions of space identified by a low density of galaxies in the LSS, are becoming increasingly important for studying the LSS and the evolution of the Universe. 
Various methods related to cosmic void for cosmological study have been discussed, such as the void abundance \citep[e.g.][]{2017MNRAS.469..787P,2019MNRAS.487.2836P,contarini2021cosmic,2022A&A...667A.162C,contarini2023cosmological,2023MNRAS.522..152P,2024MNRAS.532.1049S,2024MNRAS.534..128S,2024JCAP...10..079V,2024arXiv241019713V}, cross-correlation between void and other probes, e.g. galaxy clustering and gravitational lensing \citep[e.g.][]{2016MNRAS.462.2465C,2017MNRAS.465..746S,2018MNRAS.476.3195C,2019MNRAS.483.3472N,2021MNRAS.500..464V,2022MNRAS.516.4307W,2023A&A...670A..47B,2023JCAP...08..010V,2023arXiv231208483C,2023A&A...674A.185M,2023A&A...677A..78R,2024ApJ...976..244S,2025MNRAS.538..114S}.

As a powerful cosmological probe, cosmic void is effective for studying massive neutrinos and modified gravity \citep[e.g.][]{cai2015testing,pisani2015counting,zivick2015using,pollina2016cosmic,achitouv2016testing,sahlen2016cluster,falck2018using,sahlen2018cluster,paillas2019santiago,perico2019cosmic,verza2019void,2019MNRAS.488.4413K,2019JCAP...12..055S,2022ApJ...935..100K,2023JCAP...12..044V,2024ApJ...969...89T}. Besides, the shape of cosmic voids is suitable for testing cosmological geometrical and dynamical effects due to their large volume, such as the baryonic acoustic oscillation (BAO) \citep[e.g.][]{chan2021volume,forero2022cosmic,khoraminezhad2022cosmic,2022MNRAS.511.5492Z,2023MNRAS.526.2889T}, redshift space distortion (RSD) \citep[e.g.][]{paz2013clues,cai2016redshift,2017JCAP...07..014H,chuang2017linear,nadathur2019accurate,nadathur2020completed,correa2022redshift} and Alcock-Paczy\'{n}ski (AP) effect \citep[e.g.][]{sutter2012first,sutter2014measurement,hamaus2016constraints,mao2017cosmic,correa2021redshift,2022A&A...658A..20H,2024A&A...691A..39R}.
Some important attempts using the observational data have demonstrated the potential of cosmic voids for the cosmological study, and provided remarkable references for subsequent studies \citep[e.g.][]{hamaus2016constraints,mao2017cosmic,2017ApJ...835..161M,2020JCAP...12..023H,nadathur2020completed,2022MNRAS.513..186A,contarini2023cosmological,2024ApJ...969...89T}.


In this work, we derive the void size function (VSF), which characterizes the number density distribution of voids according to their size at a given redshift, from the sixteenth data release (DR16) of the Baryon Oscillation Spectroscopic Survey \citep[BOSS,][]{2013AJ....145...10D,2015ApJS..219...12A},
contained within the sixteenth data release of the Sloan Digital Sky Survey \citep[SDSS,][]{2006AJ....131.2332G,2020ApJS..249....3A}, which also includes data from the Extended Baryon Oscillation Spectroscopic Survey \citep[eBOSS,][]{2016AJ....151...44D}. The galaxy catalog from BOSS DR16 we use covers large contiguous sky areas and has a wide redshift coverage, which is ideal for the cosmological study using cosmic voids. We then use the VSF to investigate constraints on cosmological parameters.

The Sheth and van de Weygaert model \citep[SvdW,][]{sheth2004hierarchy} was first used to describe the VSF. It assumes voids to be spherical in shape and forms the foundation of theoretical VSF models. The SvdW model was later extended to the volume-conserving model \citep[$V{\rm d}n$,][]{jennings2013abundance}, which has become one of the most widely used VSF models. 

We identify voids in galaxy catalog from BOSS DR16 using Voronoi tessellation and the watershed algorithm. The properties of the voids, such as the volume-weighted center, effective radius, and ellipticity, are estimated. These voids are identified to have arbitrary shapes, without assuming a simple spherical form. To ensure the reliability of our results, we select the voids based on the ellipticity criterion in our VSF analysis. The VSF data are derived in two redshift bins ranging from $z=0.2$ to 0.8. We employ the Markov Chain Monte Carlo (MCMC) method to constrain the model parameters. The   theoretical model is based on the excursion-set theory, and includes the cosmological and void parameters. This method has been proven to be feasible and effective in the previous VSF analysis, which is tested using the mock data from simulations \citep{2024MNRAS.532.1049S}.

The paper is organized as follows: In Section \ref{sec:data}, we introduce the galaxy catalog of BOSS DR16 we use, and the method for generating void catalogs; In Section \ref{sec:vsf}, we discuss the calculation of the theoretical model and the estimation of the VSF data; In Section \ref{sec:mcmc}, we present the constraint results of the relevant cosmological and void parameters; The summary and conclusion are given in Section \ref{sec:conclusion}.

\section{DATA ANALYSIS} \label{sec:data}

\subsection{Galaxy catalog} \label{sec:galcat}
\begin{figure}
\centering
\includegraphics[width=\columnwidth]{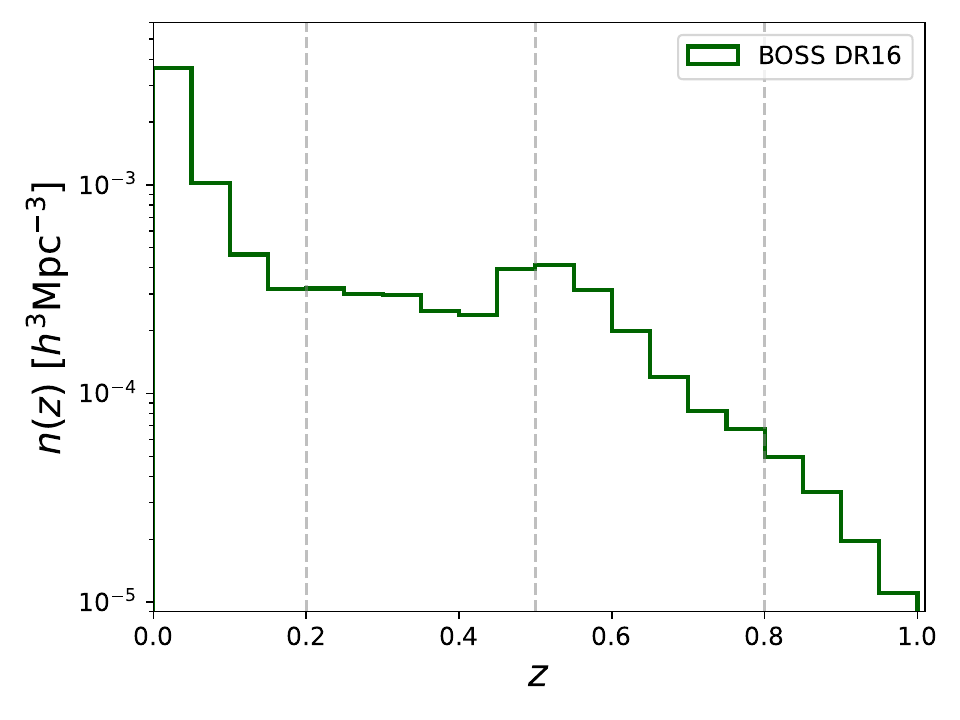}
\caption{The galaxy number density distribution of our catalog from BOSS DR16 in $0<z<1$. The redshift bins are divided by the gray vertical dashed lines.}
\label{fig:ng}
\end{figure}

We use the galaxy catalog from BOSS DR16 \citep{2013AJ....145...10D}, which covers both the Northern and Southern Galactic Caps and includes galaxies from both BOSS DR12 and eBOSS DR16 \citep[e.g. LOWZ and CMASS,][]{2017MNRAS.470.2617A,2021PhRvD.103h3533A}. This catalog covers a sky area of more than 10,000 square degrees and contains approximately four million sources with redshifts up to $z \sim 3$.
To determine the accuracy of the redshift measurement of the sources, we choose the SPectra Analysis \& Retrievable Catalog Lab\footnote{\url{https://astrosparcl.datalab.noirlab.edu/}} (SPARCL) to obtain the samples with ${\rm SPECTYPE==GALAXY}$ and ${\rm ZWARN==0}$. Here SPECTYPE is the classification of source, and ZWARN indicates the potential issues in the spectroscopic redshift measurement by a redshift fitting software $\tt Redrock$\footnote{\url{https://github.com/desihub/redrock}}. A total of 1,911,476 galaxies are retained in $ 0<z<2$. 
We assume a fiducial cosmology with $\Omega_{\rm m} = 0.3111$ and $h = 0.6766$, based on the flat $\Lambda$CDM model and the $\it Planck$ 2018 results \citep{2020A&A...641A...6P}, to convert galaxy redshifts and angular positions into comoving coordinates and compute the relevant properties of galaxy and void, such as galaxy number density, mean galaxy separation, and void ellipticity.
The galaxy number density distribution is shown in Figure~\ref{fig:ng}. 

Since the survey volume and number of galaxies are relatively small at $z<0.2$, and the galaxy number density at $z>0.8$ rapidly declines to the order of $10^{-5}$, we selected two redshift ranges, i.e. $z=0.2-0.5$ and $0.5-0.8$, based on the number density distribution in our galaxy catalog and previous studies \citep[e.g.][]{2016MNRAS.455.1553R,2020JCAP...12..023H,2022MNRAS.511.5492Z,contarini2023cosmological}. We calculate the number density of galaxies $\bar n_{\rm g}$ and the mean galaxy separation (${\rm MGS} =\bar n_{\rm g}^{-1/3}$) by the number of galaxies $N_{\rm g}$ and the survey volume $V_{\rm s}$ at the two redshift bins. 
Note that the survey volume $V_{\rm s}$ is computed in the fiducial cosmology. 
We show the $N_{\rm g}$, $\bar n_{\rm g}$ and MSG value in the two redshift bins in Table~\ref{tab:cat}.

\begin{table*}
    \caption{ The numbers and number densities of galaxy and void, i.e. $N_{\rm g}$, $N_{\rm v}$, $\bar n_{\rm g}$ and $\bar n_{\rm v}$ (in $h^3\text{Mpc}^{-3}$), in our catalog in the two redshift bins. The MGS, and minimum and maximum radius of voids (in $h^{-1}\text{Mpc}$) are also shown.}\label{tab:cat}
    \begin{tabular}{ccccccccc}
        \hline \hline
        $z_{\rm min}$ & $z_{\rm max}$&$N_{\rm g}$&$N_{\rm v}$&$\bar n_{\rm g}$&$\bar n_{\rm v}$ & MGS & $R_{\rm v}^{\text{min}}$ & $R_{\rm v}^{\text{max}}$
        \\
        \hline
        0.2&0.5&679834&3844& $3.0 \times 10^{-4}$& $1.7 \times 10^{-6}$&15&4&80\\
        0.5&0.8&963604&5934&$1.8 \times 10^{-4}$&$1.1 \times 10^{-6}$& 18 &5&113\\  
        \hline
	\end{tabular}
\end{table*}

\subsection{Void catalog} \label{sec:voidcat}

We use the Void IDentification and Examination toolkit\footnote{\url{https://bitbucket.org/cosmicvoids/vide\_public/src/master/}} \citep[\texttt{VIDE},][]{vide} to obtain the void catalog from our galaxy catalog, with galaxy coordinates converted according to the fiducial cosmology. 
\texttt{VIDE} employs Voronoi tessellation and the watershed algorithm \citep{watershed} to identify underdensity regions, based on the ZOnes Bordering On Voidness (\texttt{ZOBOV}) method \citep{zobov}. The voids identified by \texttt{VIDE} have the advantage of not assuming any specific shape, and it can provide several important void quantities, such as ellipticity, volume-weighted center, and effective radius. \texttt{VIDE} also can merge neighboring low-density regions based on the galaxy number density on the boundary, but in this work, we use the unmerged low-density zones to generate the void catalog for  avoiding the void-in-void issue.

The void volume $V$ and effective radius are calculated using the cell properties of the tracer, i.e. galaxy in our analysis. Voronoi tessellation assigns a cell to each galaxy, and the effective radius of the non-spherical void $R_{\rm v}$ is determined based on the volume of each cell $V^i_{\rm cell}$ in this void, that we have
\begin{equation}\label{eq:rv} 
R_{\rm v} = \left(\frac{3V}{4\pi} \right)^{1/3}= \left(\frac{3}{4\pi}\sum^N_{i=1} V^i_{\rm cell}\right)^{1/3}.
\end{equation}
Here $N$ is the number of cells in this void. We use the galaxy position $\mathbf{x}_i$ in each cell to calculate the volume-weighted center of a void $\mathbf{X}_{\rm v}$, which is given by
\begin{equation}
\mathbf{X}_{\rm v} = \frac{\sum^N_{i=1}\mathbf{x}_i V^i_{\rm cell}}{\sum^N_{i=1} V^i_{\rm cell}}.\label{eq:xv}
\end{equation}
On the other hand, the irregularity of the void shape can be represented by the void ellipticity $\epsilon_{\rm v}$, which can be calculated from the maximum and minimum eigenvalues of the inertia tensor, i.e., $I_{\rm max}$ and $I_{\rm min}$. Then we have 
\begin{equation}
\epsilon_{\rm v} = 1 - \left(\frac{I_{\rm min}}{I_{\rm max}}\right)^{1/4}.\label{eq:ev}
\end{equation}
The inertia tensor can be written as
\begin{equation}
I_{\rm xx} = {\sum_{i=1}^{N}}(y^2_i+z^2_i)
, \ I_{\rm xy} = -{\sum_{i=1}^{N}}x_iy_i.\label{eq:it}
\end{equation}
Here $I_{\rm xy}$ and $I_{\rm xx}$ represent the off-diagonal and diagonal components, respectively, and $x_i$, $y_i$, and $z_i$ are the galaxy coordinates in cell $i$ measured relative to the void volume-weighted center $\mathbf{X}_{\rm v}$. $I_{\rm yy}$, $I_{\rm zz}$, $I_{\rm xz}$, and $I_{\rm yz}$ can be calculated using a similar formula to Equation~(\ref{eq:it}).

\begin{figure}
\centering
\includegraphics[width=\columnwidth]{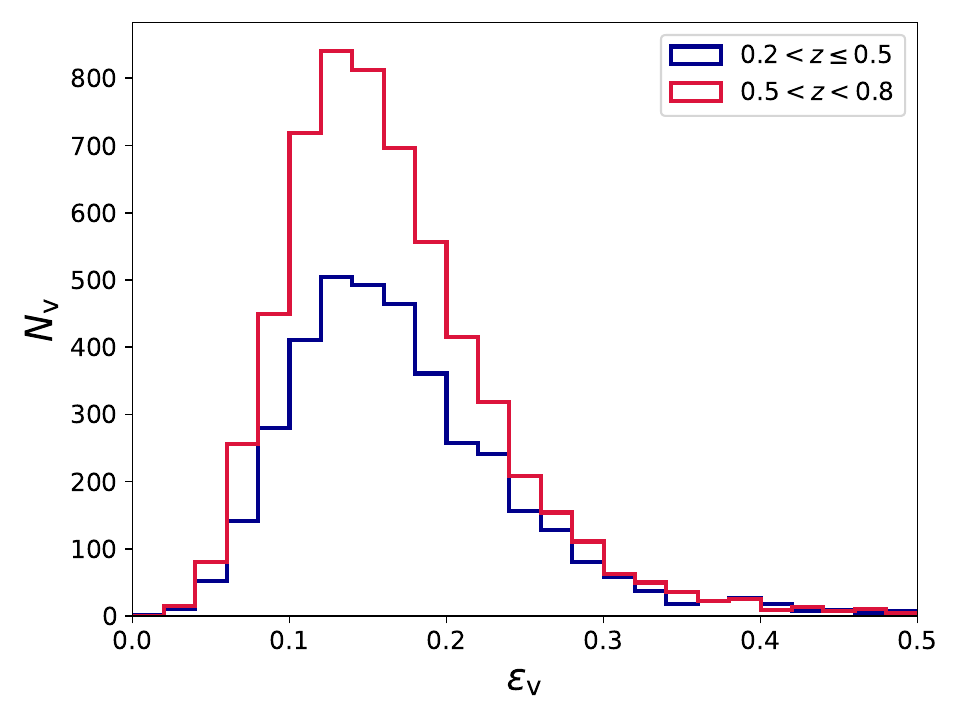}
\caption{The void ellipticity distributions of the redshift bins $z=0.2-0.5$ and $0.5-0.8$, respectively.}
\label{fig:ved}
\end{figure}

We show the void ellipticity distribution in the two redshift bins in Figure~\ref{fig:ved}. We can see that the two distributions of void ellipticity have a similar shape with a peak $\epsilon_{\rm v}^{\rm peak}\sim0.15$. 
In Table~\ref{tab:cat}, we show the number of voids $N_{\rm v}$, void number densities and the minimum and maximum void radii in our catalog. We can find that, there is no order of magnitude difference in the number densities of galaxy and void between the two redshift bins,  and the void and galaxy number densities decrease as the redshift increases, which is just as expected.

\begin{figure*}
\subfigure{
\includegraphics[width=\columnwidth]{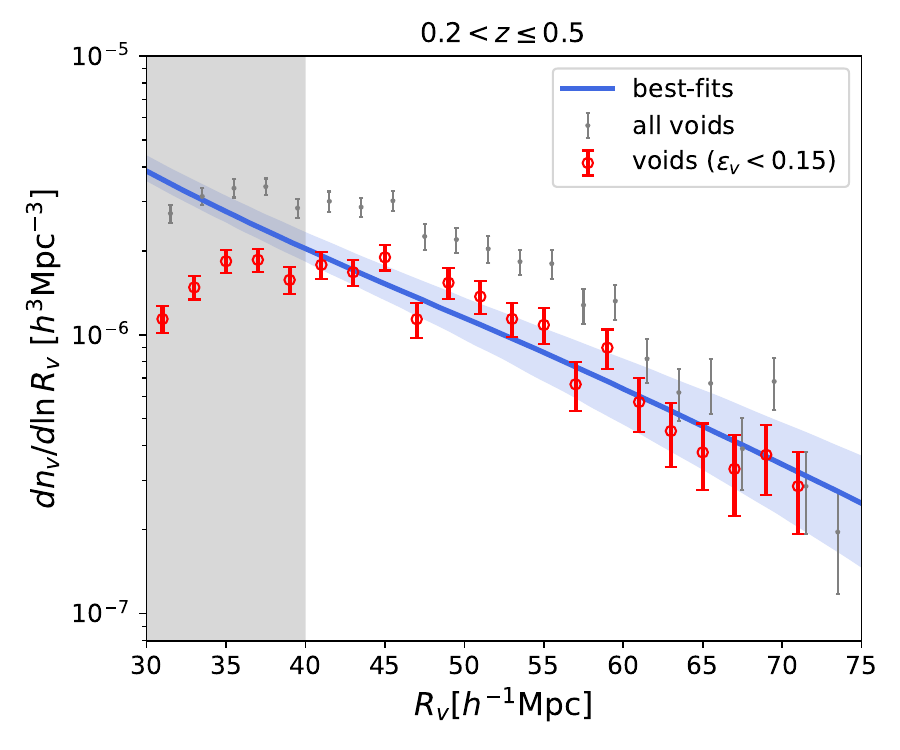}}
\hspace{2mm}
\subfigure{
\includegraphics[width=\columnwidth]{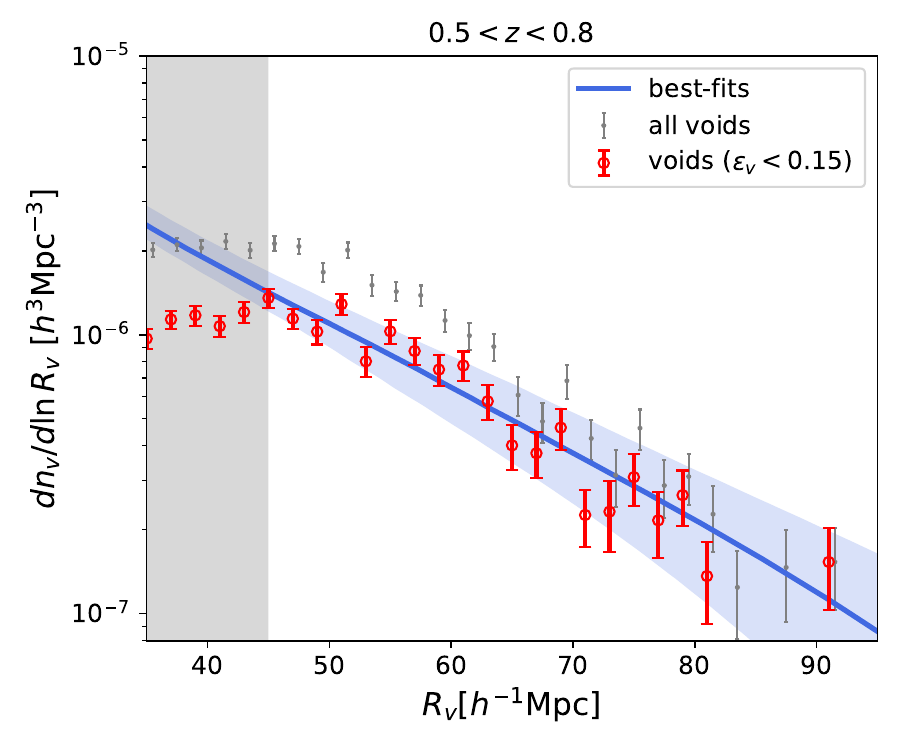}}
\caption{ The void size functions from BOSS DR16 with $\epsilon_{\rm v}<0.15$ (red) and all voids (gray) in the two redshift bins. The blue curves are the best-fits of the theoretical VSF model obtained from the MCMC process, and the blue shaded regions indicate the uncertainty of the model calculated by using the 1$\sigma$ errors of $\Omega_{\rm m}$ and $w$. The gray areas represent the excluded scales based on $\sim2.5\times$MGS in the fitting process.}\label{fig:vsf}
\end{figure*}

\section{VOID SIZE FUNCTION}
\label{sec:vsf}
The void size function we adopt is based on the volume fraction conserving model, which is known as the $V{\rm d}n$ model \citep{jennings2013abundance}. The VSF can be modeled by
\begin{equation}
\frac{{\rm d}n_{\rm v}}{{\rm d\,ln}R_{\rm v}}=(\frac{3}{4\pi R_{\rm v}^3})\mathcal{F}(\nu,\delta_{\rm v},\delta_{\rm c})\frac{d\nu}{{\rm d\,ln}R_{\rm L}}.\label{eq:vsf}
\end{equation}
Here $R_{\rm L}$ is the void radius in Lagrangian space.
In this framework, Lagrangian space refers to the initial, linearly evolved density field, which is the linear density contrast as a function of position, scaled by the appropriate linear growth factor. This represents the density distribution at sufficiently high redshift, where linear theory accurately describes its statistical properties. On the other hand, Eulerian space describes the fully non-linear, evolved density field at the same redshift.
In Equation~(\ref{eq:vsf}), $\nu$ is the significance of the void formation threshold in terms of the standard deviation of matter fluctuations \citep{2018PhR...733....1D}, which is determined by the root-mean-square density fluctuations $\sigma_M(z)$ and the void formation threshold $\delta_{\rm v}$ at different redshifts. 
It can be written as
\begin{equation}
\nu=\frac{|\delta_{\rm v}|}{\sigma_M(z)}, \label{eq:nu}    
\end{equation} 
where $\sigma_M(z)=\sigma_0(R_{\rm L})D(z)$, and $D(z)$ is the linear growth factor. We use \texttt{CAMB} to obtain $\sigma_M(z)$ in our work \citep{camb}.
$\mathcal{F}(\nu, \delta_{\rm v}, \delta_{\rm c})$ denotes the probability that a random trajectory crosses the barrier $\delta_{\rm v}$ for the first time at $\nu$ without having crossed $\delta_{\rm c}=1.686$ for any $\nu' > \nu$. It can be calculated or approximated by \citep{sheth2004hierarchy} 
\begin{equation}
\begin{split}
\mathcal{F}(\nu, \delta_{\rm v},\delta_{\rm c})&=\frac{2\mathcal{D}^2}{\nu^3}\sum^{\infty}_{j=1}j\pi\, {\rm sin}(\mathcal{D}j\pi)\,{\rm exp}(-\frac{j^2\pi^2\mathcal{D}^2}{2\nu^2})\\
&=\sqrt{\frac{2}{\pi}}\,{\rm exp}(-\frac{\nu^2}{2})\,{\rm exp}(-\frac{|\delta_\text{v}|}{\delta_\text{c}}\frac{\mathcal{D}^2}{4\nu^2}-2\frac{\mathcal{D}^4}{\nu^4}).
\end{split}
\label{eq:fnu}
\end{equation}
Here $\mathcal{D}=|\delta_{\rm v}|/(\delta_{\rm c}+|\delta_{\rm v}|)$.
Note that, in early VSF studies, the theoretical value of $\delta_{\rm v}$ was considered a fixed threshold, similar to the formation of halos \citep{jennings2013abundance}. However, more recent studies have shown that there is no strict theoretical threshold for void formation, and any negative value can be used, since voids are underdense regions defined by arbitrary density criteria \citep{pisani2015counting,2015MNRAS.454.2228N,verza2019void,2022A&A...667A.162C,2024JCAP...10..079V}.
Our voids are identified as low density regions with arbitrary shapes from our galaxy map using \texttt{VIDE}. Therefore, we take $\delta_{\rm v}^i$ as a free parameter in the fitting process in the $i$th redshift bin to match the measured VSF data.

The mapping between the Lagrangian void size $R_{\rm L}$ and the Eulerian void size $R_{\rm v}$ can be derived based on the non-linear density contrast $\delta_{\rm v}^{\rm NL}$, which reflects the transition from the linear to the non-linear regime. It is given by
\begin{equation}
R_{\rm L}=R_{\rm v} \left[ \frac{\rho}{\rho_{\rm v}}\right ] ^{-1/3}=\frac{R_{\rm v}}{(1+\delta_{\rm v}^{\rm NL})^{-1/3}},
\end{equation}
where $\rho$ is the background matter density and $\rho_{\rm v}$ is the average density in a void.
The relation between $R_{\rm v}$ and $R_{\rm L}$ can be estimated by
\begin{equation}
R_{\rm L}\simeq\frac{R_{\rm v}}{(1-\delta_{\rm v}/c_{\rm v})^{c_{\rm v}/3}},\label{eq:rvrl}
\end{equation}
where $c_{\rm v} =1.594$ and it represents the empirical best fit for the Lagrangian-to-Eulerian mapping in spherical symmetry \citep{bernardeau1993nonlinear,jennings2013abundance}.

Since the volume and effective radius of voids identified in our catalog assuming the fiducial cosmology may deviate from that in the real observations, 
we need to consider the impact of redshift space distortions (RSD) and the Alcock-Paczynski (AP) effect \citep{1979Natur.281..358A} in our analysis. The RSD effect causes voids to appear larger due to the stretching along the line of sight \citep{2015arXiv150607982P,2016MNRAS.459.2670Z,2016MNRAS.461..358N,correa2021redshift}. The AP effect introduces an anisotropy in the void shape, distorting the voids in both the parallel and perpendicular directions to the line of sight, since the true cosmology may deviate from the assumed fiducial cosmology.
We use two factors, $q_{\rm RSD}$ and $q_{\rm AP}$, to account for the corrections to the void radius from the RSD and AP effects, respectively, which are given by
\begin{equation}\label{eq:rvc}
R_{\rm v}^{\rm obs} =  q_{\rm RSD}q_{\rm AP}R_{\rm v}.
\end{equation}
The AP effect is usually considered by applying scaling factors in both the radial and transverse directions \citep{correa2021redshift}, and we have
\begin{equation}\label{eq:ap}
\alpha_{\parallel} = \frac{H^{\rm fid}(z)}{H(z)},\  \ 
{\rm and}\
\alpha_{\perp} = \frac{D_{\rm A}(z)}{ D_{\rm A}^{\rm fid}(z)}. 
\end{equation}
Here $H(z)$ and $D_{\rm A}(z)$ represent the Hubble parameter and angular diameter distance, respectively, and the superscript ``fid'' denotes the values from the fiducial cosmology. We apply these two factors to correct the void radius affected by the AP effect, where $q_{\rm AP} = \alpha_{\parallel}^{1/3} \alpha_{\perp}^{2/3}$ \citep{2017MNRAS.464.1640S,2020JCAP...12..023H,correa2021redshift,contarini2023cosmological}.

In our void sample used for analyzing the VSFs, the factor of RSD can be derived by \citep{correa2021redshift}
\begin{equation}\label{eq:rsd}
q_{\rm RSD} = 1 - \frac{1}{3}\delta R_{\rm v}\beta\Delta(R_{\rm v}),
\end{equation}
where $\beta$ is the ratio of the growth rate $f$ to the galaxy bias $b_{\rm g}$ and $\delta R_{\rm v}$ is an additional factor that quantifies
the variation in void radius, which is expected to range between 0 and 1 \citep{correa2021redshift}. We set $B^i=\beta^i\delta R_{\rm v}^i$ as a free parameter in the $i$th redshift bin in the constraints. 
The value of $\Delta(R_{\rm v})$ can be computed using the formula $\Delta(R_{\rm v}) = \delta_{\rm v}^{\rm NL} = (1-\delta_{\rm v}/c_{\rm v})^{-c_{\rm v}}-1$ \citep{bernardeau1993nonlinear}.

Since the theoretical model of the VSF is based on the spherical evolution, and the voids identified by \texttt{VIDE} using the watershed algorithm have no assumed shape, we select spherical-like voids with the void ellipticity $\epsilon_{\rm v} < 0.15$ in the fitting process. This selection criterion of the void ellipticity value is around the peak of the void ellipticity distributions in the two redshift bins as shown in Figure~\ref{fig:ved}. We find the number of voids is reduced to around 45\% for two redshift bins by applying this cut-off.

We should note that, in this analysis, since the void linear threshold $\delta_{\rm v}$ is treated as a free parameter, it allows more flexibility in modeling the VSF. Moreover, the relation between the theoretical model based on the excursion-set model of Jennings et al. (2013) and the abundance derived from the observed watershed void catalog is empirical. A physically motivated theoretical model needs to be constructed to describe the watershed voids more precisely, which will be explored in the future.


We show the void size functions for all voids and voids with $\epsilon_{\rm v}<0.15$ of the two redshift bins in Figure~\ref{fig:vsf}. We notice that, since  the ellipticity of small-size voids is relatively high \citep{2024MNRAS.532.1049S}, the difference between the theoretical model and data at small scales generally cannot be ignored. Previous studies usually exclude the small-size voids based on the value of MGS \citep[e.g.][]{2019MNRAS.488.5075R,2019MNRAS.488.3526C,contarini2021cosmic,2022A&A...667A.162C,contarini2023cosmological,2024A&A...682A..20C}. We follow this method and exclude voids with $R_{\rm v} < 40 h^{-1}\text{Mpc}$ in $0.2 < z \leq 0.5$ and $R_{\rm v} < 45 h^{-1}\text{Mpc}$ in $0.5 < z < 0.8$, which are approximately 2.5 times of the MGS in the two redshift bins. Furthermore, if the number of voids is less than 5  in a void radius bin, we exclude them in the fitting process to ensure the statistical significance.

To test the robustness of our selection criteria, we conducted tests on the ellipticity cut for each redshift bin. We find that for ellipticity thresholds of $\epsilon_{\rm v}= 0.14$ or 0.16, the constraint result would not change significantly and is consistent with the case using $\epsilon_{\rm v}= 0.15$. However, if choosing a threshold smaller than 0.14 or larger than 0.16, it leads to deviations of more than 1$\sigma$ in some parameters (e.g. $\Omega_{\rm m}$). This suggests that moderate change of the ellipticity cut is acceptable in our method, and the result is not very sensitive to this cut.
And we also find that our selection includes all approximately spherical large-size voids with ellipticity below the mean vlaue of the sample. Therefore, we consider that our current ellipticity selection criterion, which is based on the peak value from the void ellipticity distribution, provides a reasonable choice.

Note that the trimming on void ellipticity is empirically motivated. There is currently no theoretical framework that can provide an optimal ellipticity threshold to select voids corresponding to the assumptions of the \citet{jennings2013abundance} model. Based on the analysis in this work and our previous work  \citep{2024MNRAS.532.1049S}, we notice that the void ellipticity distribution is size-dependent, that larger voids tend to be more spherical while smaller voids are more flattened. This trend suggests that the future theoretical models could include size-dependent terms for helping introducing priors on the ellipticity choice to improve the robustness of cosmological constraints.

In addition, we also test different minimum void radius selections in each redshift bin to evaluate their impact on the constraints. We firstly examine two choices, i.e. one with a radius 5 $h^{-1}\text{Mpc}$ smaller ($\sim$2.3$\times$MGS) and another with a 10 $h^{-1}\text{Mpc}$ smaller radius ($\sim$2$\times$MGS) than our current choice of $\sim$2.5$\times$MGS. We find that both of these choices have significant impact on the constraints, that the results deviate by more than 1$\sigma$ for some parameters, e.g. $\Omega_{\rm m}$, $w$ and $\delta_{\rm v}$, compared to the results from the case with $\sim$2.5$\times$MGS. On the other hand, if we choose larger void radius cuts, the best-fit results of the model parameters would not change significantly, but the errors will become larger, since less data points are used in the fitting process in this case. Based on these tests, we consider that our current selection of the minimum void radius ($\sim$2.5$\times$MGS) is reasonable and can provide reliable constraint results. Besides, this minimum void radius is also consistent with the selection used in previous studies \citep[e.g.][]{2019MNRAS.488.5075R,2019MNRAS.488.3526C,contarini2021cosmic,contarini2023cosmological,2024A&A...682A..20C}.

In order to estimate the uncertainty in the VSF measurements, we calculate the covariance matrix using the jackknife method. 
The advantage of the jackknife method is that it estimates the statistical uncertainty directly from the observational catalog, using the same assumed cosmology for both the measured void size function and its covariance. Although the measurement still depends on the fiducial cosmology used to convert redshifts into distances, the method ensures consistency between the measured VSF and its error estimate.
In addition, the covariance matrix estimated using the jackknife method converges to the true covariance matrix in the limit of large volume  \citep{2021MNRAS.505.5833F}. 

\begin{figure}
\includegraphics[width=\columnwidth]{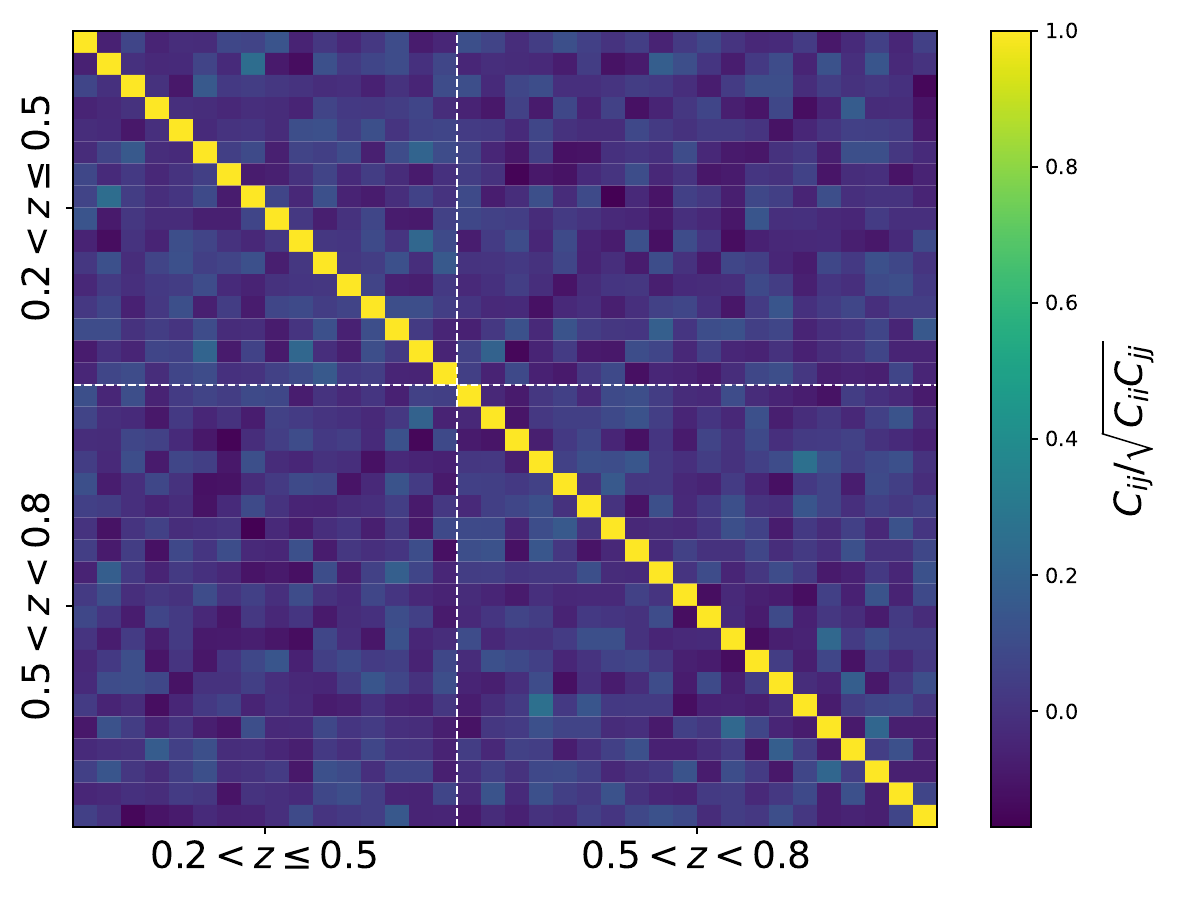}
\caption{The normalized covariance matrix of the void size function by its diagonal components. The two redshift bins are separated by the vertical and horizontal white dashed lines.}\label{fig:vsfcov}
\end{figure}

We apply the method of equally dividing the sky area to generate the jackknife subsamples for a more accurate error estimate \citep{2021MNRAS.501.3309Z}, and use 200 subsamples to compute the covariance matrix. 
In Figure~\ref{fig:vsf}, we compare the measured void size functions from BOSS DR16 with the best-fit model derived from the MCMC sampling. The uncertainty of the best-fit curve derived from the 1$\sigma$ errors of $\Omega_{\rm m}$ and $w$ is also shown in shaded blue region in each redshift bin.
The error bars for each data point in Figure~\ref{fig:vsf} are calculated from the square root of the diagonal elements of the covariance matrix. Figure~\ref{fig:vsfcov} shows the covariance matrix $C_{ij}$ normalized by its diagonal components $\sqrt{C_{ii}C_{jj}}$.

\section{CONSTRAINT AND RESULTS}
\label{sec:mcmc}

\begin{table}
    \caption{The priors and constraint results of the cosmological parameters, the void linear underdensity threshold parameter $\delta_{\rm v}^i$ and RSD parameter $B^i$ in the two redshift bins. Note that $\mathcal{U}(a,b)$ denotes a uniform prior with the range $(a,b)$, and $\mathcal{N}(\mu,\sigma)$ indicates a Gaussian prior, where $\mu$ and $\sigma$ are the mean value and standard deviation, respectively.\label{tab:mcmc}}
    \renewcommand{\arraystretch}{1.5}
    \centering
    \begin{tabular}{ccc}
    \hline\hline
    Parameter & Prior & Constraint Results\\
    \hline  
    Cosmology\\
    \hline
    $w$& $\mathcal{U}(-2.0, 0)$& $-1.263_{-0.396}^{+0.329}$\\
    $\Omega_{\rm m}$&$\mathcal{U}(0.1, 0.5)$& $0.293_{-0.053}^{+0.060}$\\
    $\sigma_8$&-& $0.897_{-0.192}^{+0.159}$\\
    $A_{\rm s}(\times 10^{-9})$&$\mathcal{N}(2.105, 0.3)$& $2.181_{-0.297}^{+0.293}$\\
    $h$&$\mathcal{N}(0.6766, 0.042)$&  $0.711_{-0.032}^{+0.033}$\\
    $\Omega_{\rm b}$&$\mathcal{N}(0.049, 0.003)$ &$0.049\pm 0.003$\\
    $n_{\rm s}$&$\mathcal{N}(0.9665, 0.038)$& $0.971_{-0.038}^{+0.037}$\\
    \hline
    Void\\
    \hline
    $\delta_{\rm v}^1$&$\mathcal{U}(-2.0, 0)$& $-0.162_{-0.024}^{+0.022}$\\
    $\delta_{\rm v}^2$&$\mathcal{U}(-2.0, 0)$& $-0.123_{-0.015}^{+0.014}$\\
    \hline
    RSD\\
    \hline
    $B^1$&$\mathcal{U}(0, 1.0)$& $0.777_{-0.267}^{+0.160}$\\
    $B^2$&$\mathcal{U}(0, 1.0)$& $0.289_{-0.209}^{+0.335}$\\
    \hline
    \end{tabular}
    \label{tab:3}
\end{table}

We adopt the standard likelihood function $\mathcal{L}$ $\propto$ exp($-\chi^2$/2) for the model fitting, and the $\chi^2$ is given by
\begin{equation}\label{eq:chi2}
\chi^2 = \left[n^{\rm obs}_{\rm v}(R_{\rm v})-n^{\rm th}_{\rm v}(R_{\rm v})\right]^{\rm T}\mathbf{C}^{-1}\left[n^{\rm obs}_{\rm v}(R_{\rm v})-n^{\rm th}_{\rm v}(R_{\rm v})\right],
\end{equation}
where $n^{\rm obs}_{\rm v}(R_{\rm v})$ is the measured VSF data, $n^{\rm th}_{\rm v}(R_{\rm v})={\rm d}n_{\rm v}/{\rm d\,ln}R_{\rm v}$ is the modeled VSF as shown in Equation (\ref{eq:vsf}), and $\mathbf{C}$ is the covariance matrix of the measured VSF. In our fitting process, we use the $n_{\rm v}(R_{\rm v})$ from the two redshift bins to obtain the joint constraint result.
We employ the MCMC method by applying \texttt{emcee} \citep{emcee,goodman} to constrain the free parameters in our model. We obtain over 1 million chain points, and remove the first 10\% of the steps as the burn-in process. Then we randomly select 10,000 points  for thinning the chains to illustrate the probability distribution functions (PDFs) of the free parameters. 

\begin{figure*}
\subfigure{
\includegraphics[width=\columnwidth]{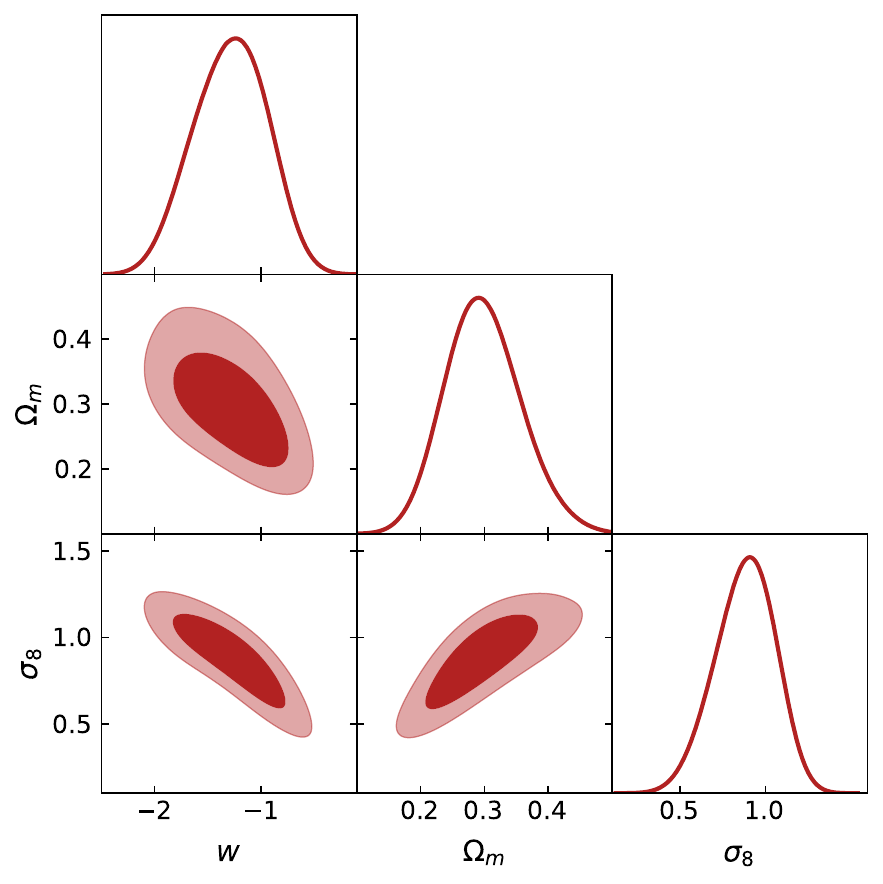}}
\hspace{1mm}
\subfigure{
\includegraphics[width=\columnwidth]{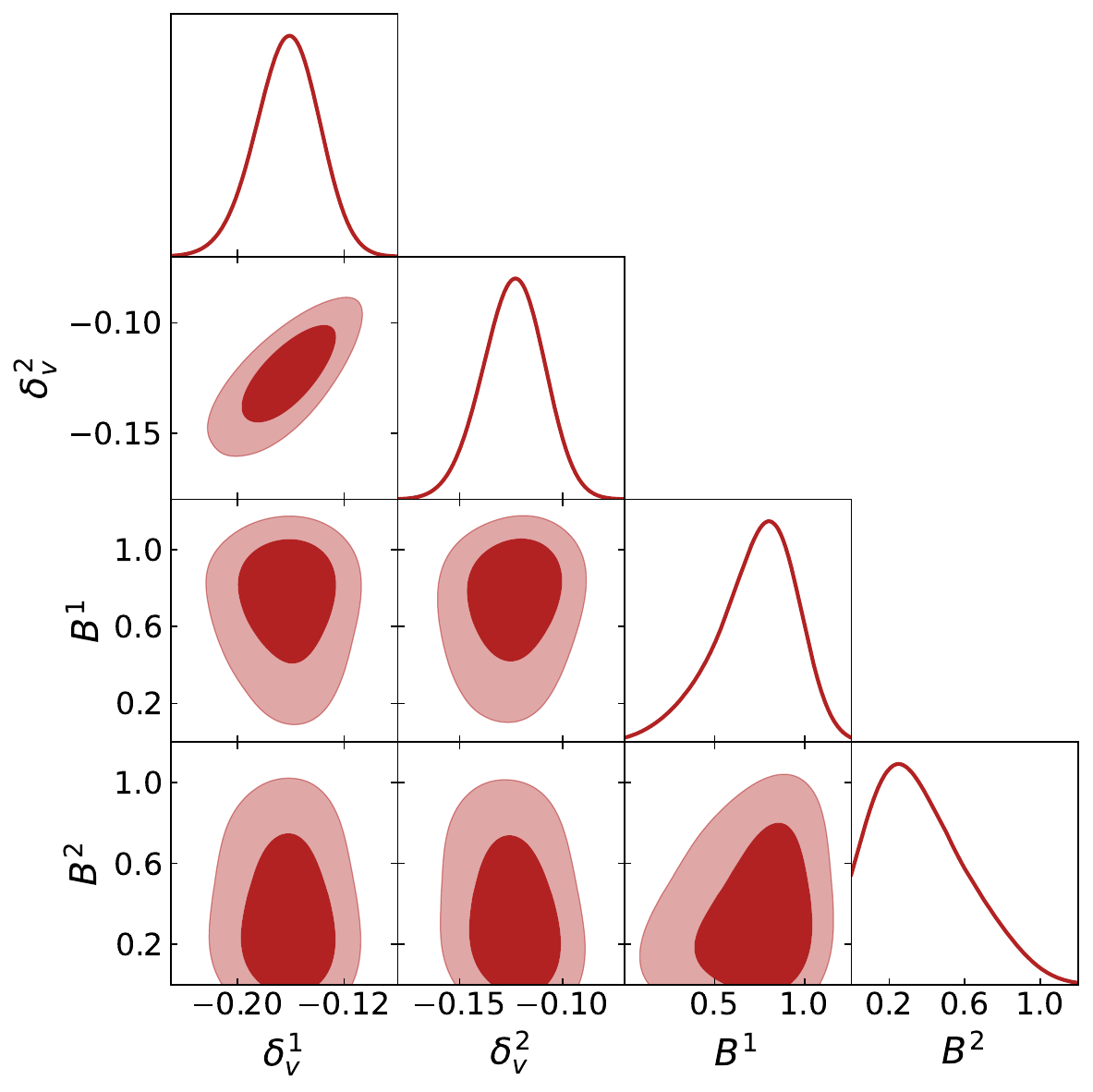}}
\caption{
The 1D PDFs and contour maps at 68\% and 95\% CL for $w$, $\Omega_{\rm m}$, and $\sigma_8$ (left panel), and $\delta_{\rm v}^i$ and $B^i$ in the two redshift bins (right panel), using the VSF data from BOSS DR16 data.}\label{fig:mcmccosmic}
\end{figure*}

In our analysis, we adopt the $w$-cold dark matter ($w$CDM) model and focus on the total matter density parameter $\Omega_{\rm m}$, the dark energy equation of state $w$ and the amplitude of matter density fluctuations on a scale of 8 $h^{-1}\text{Mpc}$, $\sigma_8$. To improve the constraints on the three parameters we are interested in, we apply a 10$\sigma$ Gaussian prior to the remaining cosmological parameters, such as the reduced Hubble constant $h$, baryon density parameter $\Omega_{\rm b}$, and spectral index $n_{\rm s}$, the amplitude of the initial power spectrum $A_{\rm s}$, based on the results from \textit{Planck}2018 \citep{2020A&A...641A...6P}.  These priors are reasonable and consistent with other observations and analyses \citep[e.g.][]{2015JCAP...07..011A,2018ApJ...855..102C,2019JCAP...10..029S,2017MNRAS.470.2617A,2021PhRvD.103h3533A,2022PhRvD.105d3517P,contarini2023cosmological}. 
Note that we convert $A_{\rm s}$ and derive the constraint result of $\sigma_8$ from the chain points.

In addition to these cosmological parameters, we also consider four other parameters, including the void linear underdensity threshold parameter $\delta_{\rm v}$ and the RSD parameter $B$ in the two redshift bins. We constrain all the parameters simultaneously in the fitting process. The priors and constraint results for all the free parameters are shown in Table~\ref{tab:mcmc}. 


In the left panel of Figure~\ref{fig:mcmccosmic}, we show the 1D PDFs and contour maps at 68\% and 95\% confidence level (CL) of $\Omega_{\rm m}$, $w$ and $\sigma_8$. 
We can find that the cosmological parameters are well constrained with accuracies $w\sim29\%$, $\Omega_{\rm m}\sim19\%$, $\sigma_8\sim20\%$, and the 1$\sigma$ CLs are consistent with the previous results from BOSS DR12 \citep[e.g.][]{contarini2023cosmological}. 
We notice that the constraints on $w$ and $\Omega_{\rm m}$ are comparable to those obtained using the VSFs from BOSS DR12, although we use the non-spherical void samples to fit the VSF model and have more free parameters. The constraint on $\sigma_8$ is relatively weak, due to more free parameters are considered and more flexible priors are assumed in our work.
These results indicate that our approach of selecting voids based on void ellipticity and MGS values is effective for obtaining reliable VSF data.
The constraint results for all free parameters are presented in the Appendix \ref{sec:appendix}.

We also note that the degeneracy directions of the $\Omega_{\rm m}-\sigma_8$, $\Omega_{\rm m}-w$ and $w-\sigma_8$ contours are consistent with the previous work using mock catalogs \citep[e.g.][]{2022A&A...667A.162C}. And the degeneracy direction of $\Omega_{\rm m}-\sigma_8$ contour is almost orthogonal to that obtained from other probes, such as galaxy clustering and the halo mass function \citep[e.g.,][]{2022A&A...667A.162C,2023MNRAS.522..152P}. This suggests that the VSF measurement can be complementary to the probes for overdense regions, e.g. galaxy clustering, and could  effectively break the degeneracies of the relevant parameters in the joint analysis.

To check the effect of the 10$\sigma$ Gaussian priors of $h$, $\Omega_{\rm b}$, and $n_{\rm s}$ on the constraint results, we also consider the 1$\sigma$ \textit{Planck}2018 Gaussian priors for the three parameters. 
In this case, we find that $w=-1.081_{-0.397}^{+0.271}$, $\Omega_{\rm m}=0.323_{-0.060}^{+0.064}$ and $\sigma_8=0.873_{-0.196}^{+0.174}$. 
We can see that the constraint results of the two cases are consistent in 1$\sigma$ and the constraint accuracies are similar, although the best-fit values of $w$ and $\Omega_{\rm m}$ from the 1$\sigma$ prior case are closer to the \textit{Planck}2018 results. This means that our constraint result is not sensitive to the choice of the Gaussian priors of $h$, $\Omega_{\rm b}$, and $n_{\rm s}$. The contour maps for the 1$\sigma$ Gaussian prior case are shown in the Appendix \ref{sec:appendix}.

Our results show that the void size function is sensitive to the cosmological parameters $\Omega_{\rm m}$, $\sigma_8$, and $w$, which directly affect the amplitude and growth of matter density fluctuations. In contrast, parameters such as $n_{\rm s}$, $h$, and $\Omega_{\rm b}$ are poorly constrained by the VSF, and varying their priors has insignificant impact on the posteriors of $\Omega_{\rm m}$, $\sigma_8$, and $w$.
The weak impact arises from the fact that these parameters have limited influence on the large-scale structure at the scales probed by voids. The $\Omega_{\rm b}$ affects baryon acoustic oscillations in the matter power spectrum but has little impact on its overall shape and amplitude at void-relevant scales. The $n_{\rm s}$ mainly affects the shape of the power spectrum at very large scales (low $k$) beyond typical void sizes. The $h$ has degeneracies with other parameters, such as the free parameter $B$ in the RSD effect, which makes it poorly constrained and insensitive to the constraint results of $\Omega_{\rm m}$, $\sigma_8$, and $w$. 

In the right panel of Figure~\ref{fig:mcmccosmic}, we show the results for the linear underdensity thresholds for void formation $\delta_{\rm v}^i$ and the RSD parameter $B^i$ in the two redshift bins. The best-fit values and 1$\sigma$ errors for $\delta_{\rm v}^i$ and $B^i$ in each redshift bin are listed in Table~\ref{tab:mcmc}. 
We find that $\delta_{\rm v}^i$ is well constrained, with the accuracy smaller than $15\%$. 
The constraint on $B^i$ are weak, and the best-fit value of $B$ decreases as redshift increases, which is basically due to that the galaxy bias increases at larger redshifts.

We also find that the best-fit values of $\delta_{\rm v}$ are ranging from $\sim-0.2$ to $-0.1$ in the two redshift bins, which increase with redshift. This is consistent with the previous study using the mock VSF data from simulations \citep{2024MNRAS.532.1049S}. 
It is mainly because that our voids are identified by the watershed algorithm without assuming the spherical shape, and our void tracers are galaxies, i.e. biased objects, instead of the particles. Additionally, the increasing trend of $\delta_{\rm v}$ is due to that the void size becomes larger (as shown in Figure~\ref{fig:vsf}) and the evolution of voids is more linear at higher redshifts, which means the Eulerian void size $R_{\rm v}$ becomes closer and closer to the Lagrangian void size $R_{\rm L}$ as redshift increasing. As shown in Equation~(\ref{eq:rvrl}), this will make $\delta_{\rm v}$ closer to 0 from low to high redshifts. Besides, we notice that the best-fit values of $\delta_{\rm v}$ from BOSS DR16 are higher than that from the mock data given in \cite{2024MNRAS.532.1049S}, which is caused by the lower galaxy number density and larger void size in the BOSS DR16 data.

We also investigate the constraint on $\delta_{\rm v}^i$ and $B^i$ from the 1$\sigma$ \textit{Planck}2018 Gaussian prior case, and find that $\delta_{\rm v}^1=-0.150_{-0.018}^{+0.017}$ and $\delta_{\rm v}^2=-0.116_{-0.011}^{+0.011}$.
We can find that the constraint results of the four parameters are in good agreement for both cases. The best-fit value of $\delta_{\rm v}^i$ from the 1$\sigma$ prior case slightly increase compared to the 10$\sigma$ prior case, with a relative accuracy improvement of about 20\%, while the results for $B^i$ are similar for the two cases. The constraint results of the $\delta_{\rm v}^i$ and $B^i$ from the 1$\sigma$ prior case are shown in the Appendix \ref{sec:appendix}.

\section{Summary and conclusion}
\label{sec:conclusion}

In this work, we investigate the void size function measured from BOSS DR16, and explore the constraints on the cosmological and void parameters. We identify voids in the BOSS DR16 galaxy catalog at $0.2 < z < 0.8$ by employing Voronoi tessellation and the watershed algorithm without assuming the spherical shape. The voids with the ellipticity $\epsilon_{\rm v}<0.15$ are selected, and the small-size voids which are smaller than 2.5$\times$MGS value in a redshift bin are excluded. Then the void catalog and the VSFs in the two redshift bins are derived for the cosmological constraint. 

We utilize the MCMC method to jointly constrain the cosmological and void parameters, and assume 10$\sigma$ Gaussian priors from $\it Planck$18 results for $A_{\rm s}$, $h$, $\Omega_{\rm b}$, and $n_{\rm s}$. We obtain $w = -1.263_{-0.396}^{+0.329}$, $\Omega_{\rm m} = 0.293_{-0.053}^{+0.060}$, and $\sigma_8 = 0.897_{-0.192}^{+0.159}$, 
which are in good agreement with the results from BOSS DR12 data. Besides, we find that the degeneracy direction of $\Omega_{\rm m}-\sigma_8$ from the VSF is almost orthogonal to that from the galaxy clustering and halo mass function measurements. This indicates that the VSF can be an effective complementary probe to the  probes for the overdense regions.
In addition, the best-fit value of the void linear threshold parameter $\delta_{\rm v}$ increases and is close to 0 at higher redshifts, which is reasonable and consistent with the previous study using the similar method applied to the mock data from simulations. 

The VSF probe can be more important in the ongoing and upcoming galaxy surveys that can perform spectroscopic measurements, such as the Dark Energy Spectroscopic Instrument \citep[DESI,][]{2016arXiv161100036D}, 4-metre Multi-Object Spectroscopic Telescope \citep[4MOST,][]{2019Msngr.175....3D,2019Msngr.175...12W}, MUltiplexed Survey Telescope \citep[MUST,][]{2024arXiv241107970Z}, Nancy Grace Roman Space Telescope \citep[RST,][]{2019arXiv190205569A}, {\it Euclid} \citep{2022A&A...662A.112E}, and China Space Station Telescope  \citep[CSST,][]{zhan11,zhan2021csst,gong,2023MNRAS.519.1132M}. These surveys are expected to greatly increase the number of observed galaxies and voids by one or two orders of magnitude compared to BOSS, and the constraint accuracy of the cosmological parameters also can be significantly improved. Our VSF method also can provide a reference for the void analysis in the next-generation galaxy surveys.

\section*{Acknowledgements}
YS and YG acknowledge the support from National Key R\&D Program of China grant Nos. 2022YFF0503404, 2020SKA0110402, and the CAS Project for Young Scientists in Basic Research (No. YSBR-092). KCC acknowledges the support the National Science Foundation of China under the grant number 12273121. XLC acknowledges the support of the National Natural Science Foundation of China through Grant Nos. 12361141814, and the Chinese Academy of Science grants ZDKYYQ20200008. This work was also supported by science research grants from the China Manned Space Project with grant nos. CMS-CSST-2025-A02, CMS-CSST-2021-B01 and CMS-CSST-2021-A01.
\section*{Data Availability}

 The data that support the findings of this study are available from the corresponding author upon reasonable request.



\bibliographystyle{mnras}
\bibliography{vsfbossref} 




\appendix
\section{MCMC Results and the Effect of Priors}\label{sec:appendix}

\begin{figure*}
\centering
\includegraphics[width=2\columnwidth]{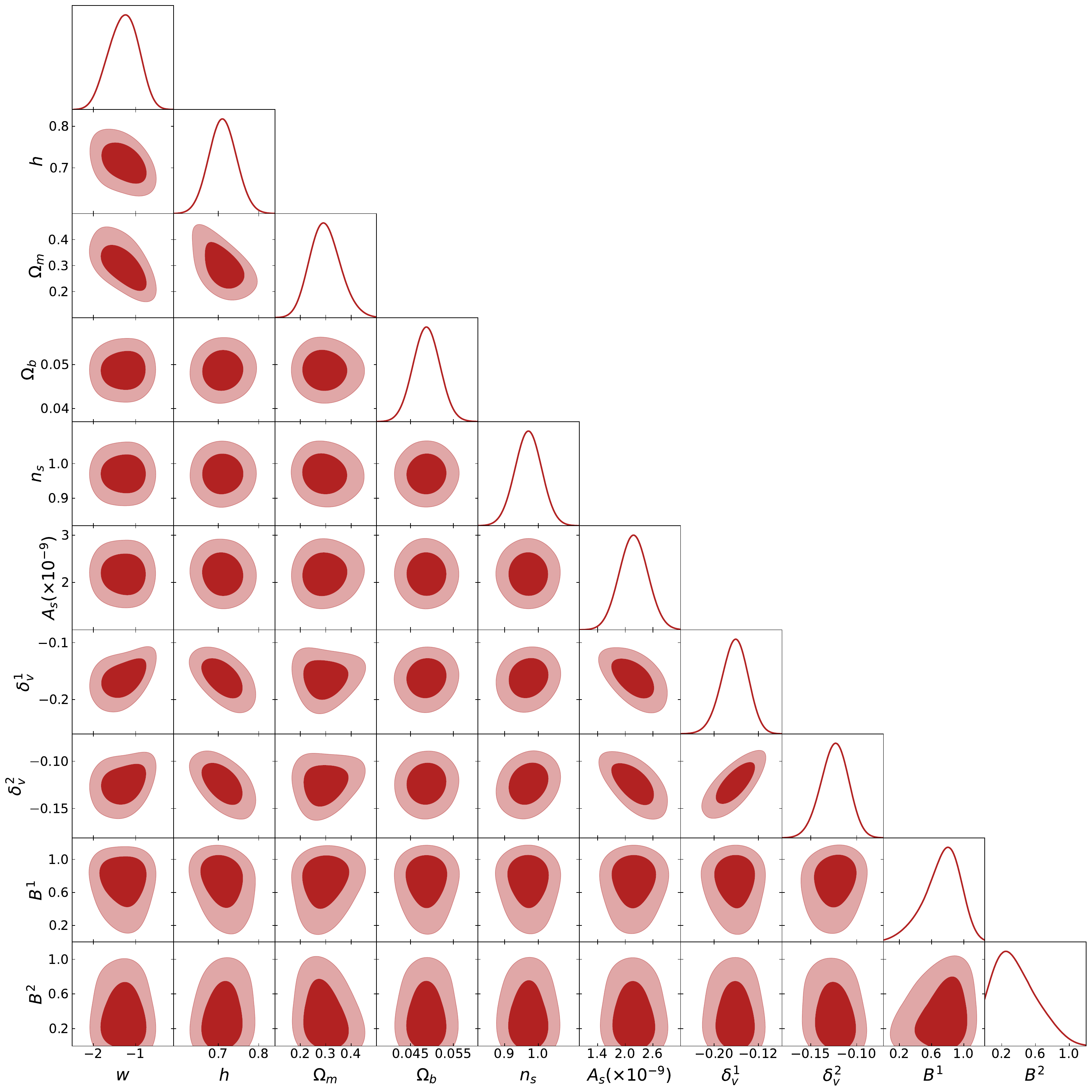}
\caption{
The 1D PDFs and contour maps at 68\% and 95\% CL for all the cosmological and void parameters in the model. The 10$\sigma$ Gaussian priors have been applied to $A_{\rm s}$, $h$, $\Omega_{\rm b}$, and $n_{\rm s}$ based on  \textit{Planck}2018 result in the fitting process.}
\label{fig:mcmcap}
\end{figure*}

In Figure~\ref{fig:mcmcap}, we show the 1D PDFs and contour maps at 68\% and 95\% CL for all the free parameters in the model. The details of the priors, best-fit values and 1$\sigma$ errors for the cosmological parameters,  as well as the linear underdensity thresholds for void formation $\delta_{\rm v}^i$ and RSD parameter $B^i$ in the two redshift bins are shown in Table \ref{tab:mcmc}. 

\begin{figure*}
\subfigure{
\includegraphics[width=1\columnwidth]{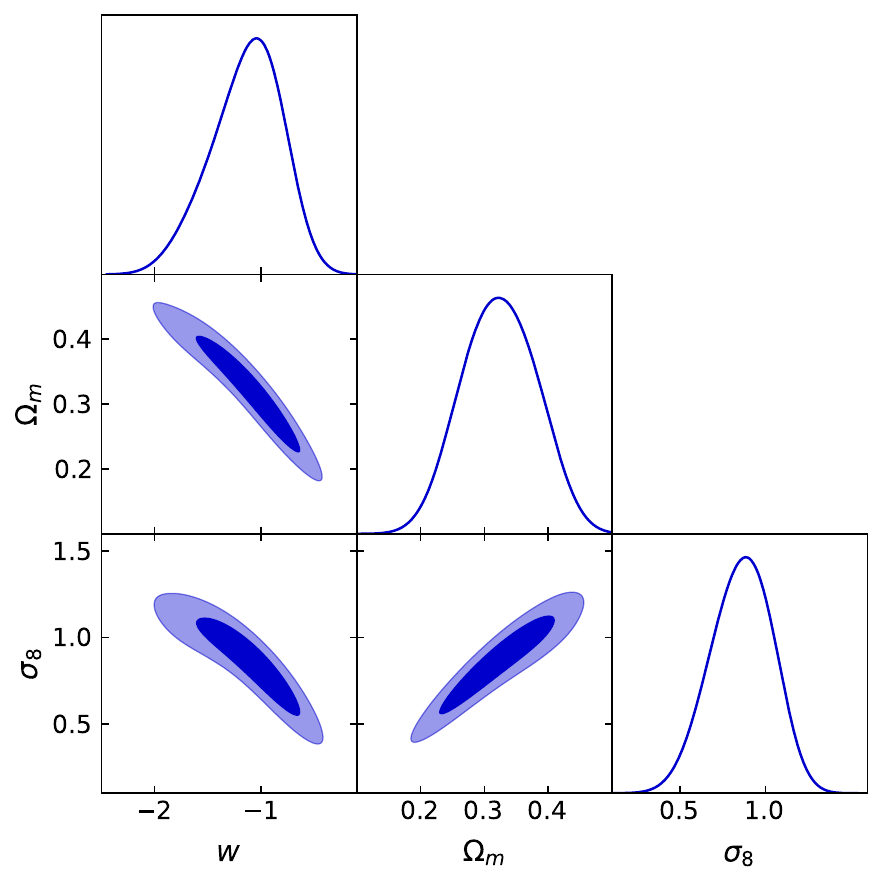}}
\hspace{1mm}
\subfigure{
\includegraphics[width=1\columnwidth]{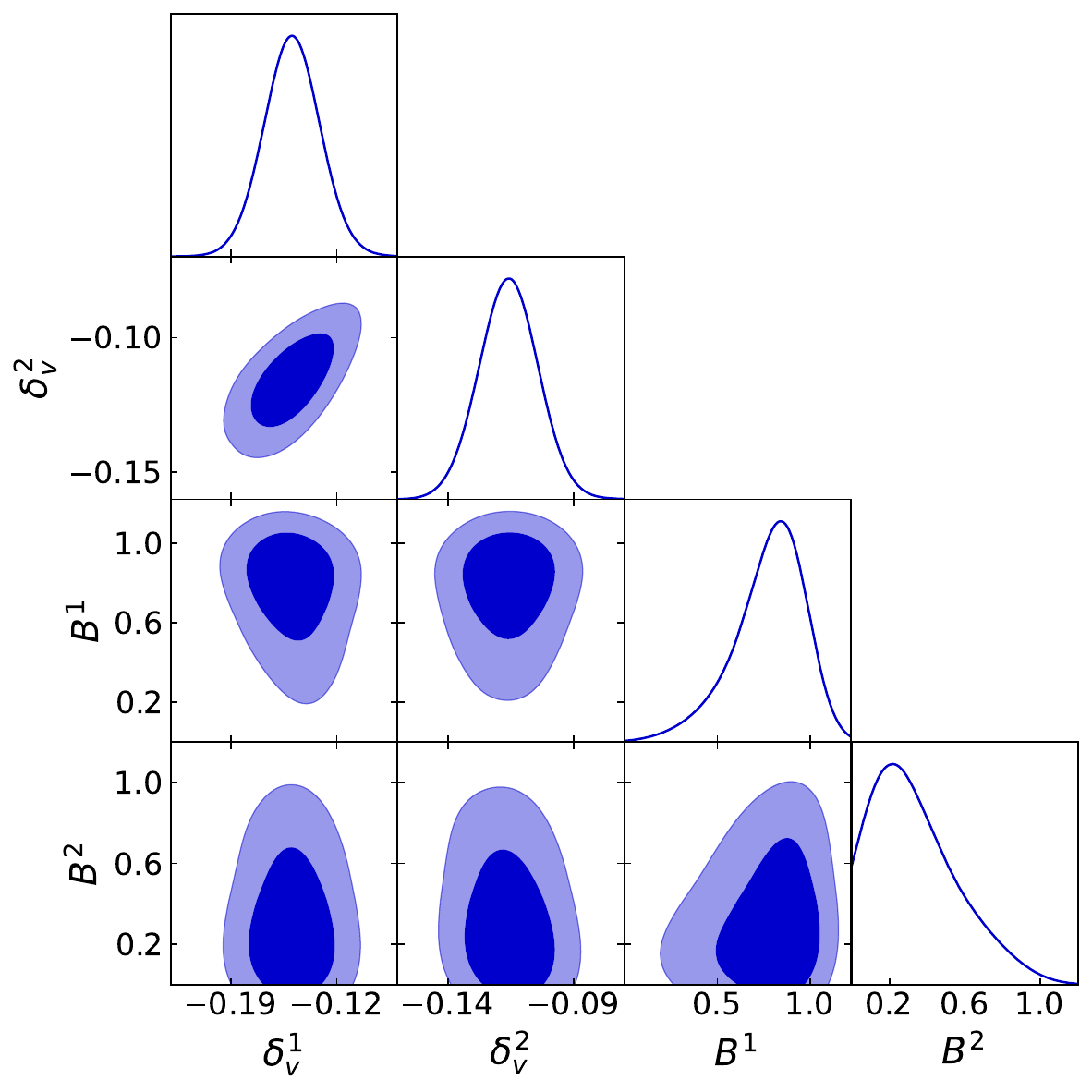}}
\caption{
The 1D PDFs and contour maps at 68\% and 95\% CL for $w$, $\Omega_{\rm m}$ and $\sigma_8$ (left panel), and $\delta_{\rm v}^i$ and $B^i$ in the two redshift bins (right panel). The 10$\sigma$ Gaussian priors have been applied to $A_{\rm s}$ and the 1$\sigma$ Gaussian priors are set to $h$, $\Omega_{\rm b}$, and $n_{\rm s}$ based on \textit{Planck}2018 result in the fitting process.}\label{fig:mcmc1scase}
\end{figure*}

In order to check the impact of the assumed Gaussian priors of $h$, $\Omega_{\rm b}$, and $n_{\rm s}$ on the cosmological constraint, we also apply stronger priors, i.e., 1$\sigma$ Gaussian priors from \textit{Planck}2018, for $h$, $\Omega_{\rm b}$ and $n_{\rm s}$. This choice is almost equivalent to fix the values of $h$, $\Omega_{\rm b}$, and $n_{\rm s}$, since the 1$\sigma$ errors from \textit{Planck}2018 are extremely small compared to that from the VSF measurement. Our results for the cosmological parameters, which include $w$, $\Omega_{\rm m}$, $\sigma_8$, and the four nuisance parameters $\delta_{\rm v}$ and $B$ from the two redshift bins are displayed in Figure~\ref{fig:mcmc1scase}.

\bsp	
\label{lastpage}
\end{document}